# ANALYTICAL SOLUTIONS OF THE SCHRODINGER EQUATION FOR HUA POTENTIAL WITHIN THE FRAMEWORK OF TWO APPROXIMATIONS SCHEME

## C. M. Ekpo[1] and E. B. Ettah[1]

[1]Department of Physics, Cross River University of Technology, Calabar

**ABSTRACT**

In this paper, we solve analytically the Schrodinger equation for s-wave and arbitrary angular momenta with the Hua potential is investigated respectively. The wave function as well as energy equation are obtained in an exact analytical manner via the Nikiforov Uvarov method using two approximations scheme. Some special cases of this potentials are also studied.

1. ## INTRODUCTION

It is well known that Bound state solutions of Schrodinger equation play an integral role in Quantum Mechanics. Bound state solutions describe the system of a particle subjected to a potential, with the tendency to remain in a fixed region of space [1]. Quantum mechanical Wavefunctions and their corresponding eigenvalues give significant information in describing various quantum systems [2-21].

Ikot et al. [20] studied the Dirac equation in the presence of the modified Mobius square potential with a Yukawa-like tensor interaction and obtain the eigenvalues and corresponding eigenfunctions for any -state by using Nikiforov-Uvarov method [20]. Ekpo *et al* [22] solved the Schrodinger equation for the New Generalized Morse-Like Potential in arbitrary dimensions by using the Nikiforov Uvarov Method. Hua potential [18], proposed by Hua as an intermolecular potential and widely applied to molecular physics and quantum chemistry.

The Hua potential is given[18];

$$V(r) = V_1 \left(\frac{1-e^{-2\alpha r}}{1-qe^{-2\alpha r}}\right)^2 \tag{1a}$$

It has been established that the solution for the Schrodinger equation for $l \neq 0$, one has to use a Pekeris-type approximation scheme to deal with the centrifugal term or solve numerically. The following new improved approximation scheme to deal with the centrifugal term (Pekeris-type approximation scheme) [14,15,16, 17]:

$$\frac{1}{r^2} = 4\alpha^2 \left(C_0 + \frac{e^{-2\alpha r}}{1-e^{-2\alpha r}}\right) = 4\alpha^2 \left(C_0 + \frac{e^{-2\alpha r}}{1-e^{-2\alpha r}} + \left(\frac{e^{-2\alpha r}}{1-e^{-2\alpha r}}\right)^2\right) \tag{1b}$$

Greene–Aldrich approximation scheme [11], which is given by

$$\frac{1}{r^2} = \frac{4\alpha^2 e^{-2\alpha r}}{(1-e^{-2\alpha r})^2} \tag{1c}$$

This paper is organized as follows. In Sect. 2, the review of the NU method is presented. In Sect. 3, this method is applied to the radial Schrodinger equation to find the analytical solution with Hua potential. In Sect. 3, the NU method is also applied to radial Schrodinger equation with Hua potential. In Sect. 4, a brief conclusion is given.

## 2. REVIEW OF NIKIFOROV-UVAROV METHOD

The Nikiforov-Uvarov (NU) method is based on solving the hypergeometric-type second-order differential equations by means of the special orthogonal functions[18]. The main equation which is closely associated with the method is given in the following form [19]

$$\psi''(s) + \frac{\tilde{\tau}(s)}{\sigma(s)}\psi'(s) + \frac{\tilde{\sigma}(s)}{\sigma^2(s)}\psi(s) = 0 \tag{2}$$

Where $\sigma(s)$ and $\tilde{\sigma}(s)$ are polynomials at most second-degree, $\tilde{\tau}(s)$ is a first-degree polynomial and $\psi(s)$ is a function of the hypergeometric-type.

The exact solution of Eq. (2) can be obtained by using the transformation

$$\psi(s) = \phi(s)y(s) \tag{3}$$

This transformation reduces Eq. (2) into a hypergeometric-type equation of the form

$$\sigma(s)y''(s) + \tau(s)y'(s) + \lambda y(s) = 0 \tag{4}$$

The function $\phi(s)$ can be defined as the logarithm derivative

$$\frac{\phi'(s)}{\phi(s)} = \frac{\pi(s)}{\sigma(s)} \tag{5}$$

where $\pi(s) = \frac{1}{2}[\tau(s) - \tilde{\tau}(s)]$ 
(5a)

with $\pi(s)$ being at most a first-degree polynomial. The second $\psi(s)$ being $y_n(n)$ in Eq. (3), is the hypergeometric function with its polynomial solution given by Rodrigues relation

$$y^{(n)}(s) = \frac{B_n}{\rho(s)} \frac{d^n}{ds^n}[\sigma^n \rho(s)] \tag{6}$$

Here, $B_n$ is the normalization constant and $\rho(s)$ is the weight function which must satisfy the condition

$$(\sigma(s)\rho(s))' = \sigma(s)\tau(s) \tag{7}$$

$$\tau(s) = \tilde{\tau}(s) + 2\pi(s) \tag{8}$$

It should be noted that the derivative of $\tau(s)$ with respect to $s$ should be negative. The wavefunction and energyequation can be obtained using the definition of the following function $\pi(s)$ and parameter $\lambda$, respectively:

$$\pi(s) = \frac{\sigma'(s) - \tilde{\tau}(s)}{2} \pm \sqrt{\left(\frac{\sigma'(s) - \tilde{\tau}(s)}{2}\right)^2 - \tilde{\sigma}(s) + k\sigma(s)} \tag{9}$$

where $k = \lambda - \pi'(s)$ \hfill (10)

The value of $k$ can be obtained by setting the discriminant of the square root in Eq. (9) equal to zero. As such, the new eigenvalue equation can be given as

$$\lambda_n = -n\tau'(s) - \frac{n(n-1)}{2}\sigma''(s), n = 0,1,2, \dots \tag{11}$$

### 3. Bound State Solution with Hua Potential

The time independent Schrodinger equation can be given as:

$$\frac{d^2 R_{nl}}{dr^2} + \frac{2\mu}{\hbar^2}\left[E_{nl} - V(r) - \frac{\hbar^2 \ell(\ell+1)}{2\mu r^2}\right] R_{nl} = 0 \tag{12}$$

where $\mu$ is the reduced mass, $E_{nl}$ is the energy spectrum, $\hbar$ is the reduced Planck's constant and $n$ and $l$ are the radial and orbital angular momentum quantum numbers respectively (or vibration-rotation quantum number in quantum chemistry). Substituting equation (1a) into

Again substituting equation (1a) into equation (12) gives:

$$\frac{d^2 R_{nl}}{dr^2} + \frac{2\mu}{\hbar^2}\left[E_{nl} - V_1\left(\frac{1-e^{-2\alpha r}}{1-qe^{-2\alpha r}}\right)^2 - \frac{\hbar^2 \ell(\ell+1)}{2\mu r^2}\right] R_{nl} = 0 \tag{13}$$

Substituting Approximation 2(Equation 1e) and $q - deforming$ Eq.1c, equation 13 becomes;

$$\frac{d^2 R_{nl}}{dr^2} + \frac{2\mu}{\hbar^2}\left[E_{nl} - V_1\left(\frac{1-e^{-2\alpha r}}{1-qe^{-2\alpha r}}\right) - \frac{\hbar^2 \ell(\ell+1)}{2\mu}\left(\frac{4\alpha^2 e^{-2\alpha r}}{(1-qe^{-2\alpha r})^2}\right)\right] R_{nl} = 0 \tag{14}$$

Rearranging Eq. 14 we get;

$$\frac{d^2 R_{n\ell}(r)}{dr^2} + \frac{2\mu}{\hbar^2(1-qe^{-2\alpha r})^2}\left[E_{nl}(1-qe^{-2\alpha r})^2 - V_1(1-e^{-2\alpha r})^2 - \frac{2\alpha^2 e^{-2\alpha r}\hbar^2 \ell(\ell+1)}{\mu}\right] R_{n\ell}(r) = 0$$
(15)

Again invoking the transformation $s = e^{-\alpha r}$ in Equation 15 we get;

$$\frac{d^2 R_{n\ell}(s)}{ds^2} + \frac{1-qs}{s(1-qs)}\frac{dR_{n\ell}(s)}{ds} + \frac{1}{s^2(1-qs)^2}[-s^2(\varepsilon_n q^2 + \beta) + s(2\varepsilon_n q + 2\beta - \gamma) - (\varepsilon_n + \beta)]R_{n\ell}(s) = 0 \qquad (16)$$

where

$$\left\{-\varepsilon_n = \frac{\mu E_{n\ell}}{2\hbar^2 \alpha^2}, \; \beta = \frac{\mu V_1}{2\hbar^2 \alpha^2} \; and \; \gamma = \ell(\ell+1)\right\} \qquad (17)$$

Comparing Eq. (16) and Eq. (2), we have the following parameters

$$\begin{cases} \tilde{\tau}(s) = 1 - qs \\ \sigma(s) = s(1 - qs) \\ \tilde{\sigma}(s) = -s^2(\varepsilon_n q^2 + \beta) + s(2\varepsilon_n q + 2\beta - \gamma) - (\varepsilon_n + \beta) \end{cases} \qquad (18)$$

Substituting these polynomials into Eq. (9), we get $\pi(s)$ to be

$$\pi(s) = -\frac{qs}{2} \pm \sqrt{\left(\frac{q^2}{4} + \varepsilon_n q^2 + \beta - kq\right)s^2 + (-2\varepsilon_n q - 2\beta + \gamma + k)s + \varepsilon_n + \beta} \qquad (19)$$

To find the constant $k$, the discriminant of the expression under the square root of Eq. (19) must be equal to zero. As such, we have that

$$k_\pm = (2\beta - \gamma - 2\beta q) \pm 2\sqrt{\varepsilon_n + \beta}\sqrt{\left(\frac{q^2}{4} - 2\beta q + \beta q^2 + \beta + \gamma q\right)} \qquad (20)$$

Substituting Eq. (20) into Eq. (19) yields

$$\pi(s) = -\frac{qs}{2} \pm \left[\left(\sqrt{\frac{q^2}{4} - 2\beta q + \beta q^2 + \beta + \gamma q} + q\sqrt{\varepsilon_n + \beta}\right)s - \sqrt{\varepsilon_n + \beta}\right] \qquad (21)$$

Again from the knowledge of NU method, we choose the expression $\pi(s)_-$ which the function $\tau(s)$ has a negative derivative. This is given by

$$k_- = (2\beta - \gamma - 2\beta q) - 2\sqrt{\varepsilon_n + \beta}\sqrt{\left(\frac{q^2}{4} - 2\beta q + \beta q^2 + \beta + \gamma q\right)} \qquad (22)$$

with $\tau(s)$ being obtained as

$$\tau(s) = 1 - 2qs - 2\left[\left(\sqrt{\frac{q^2}{4} - 2\beta q + \beta q^2 + \beta + \gamma q} + q\sqrt{(\varepsilon_n + \beta)}\right)s - \sqrt{(\varepsilon_n + \beta)}\right] \qquad (23)$$

Referring to Eq. (10), we define the constant $\lambda$ as

$$\lambda = (2\beta - \gamma - 2\beta q) - 2\sqrt{\varepsilon_n + \beta}\sqrt{\left(\frac{q^2}{4} - 2\beta q + \beta q^2 + \beta + \gamma q\right)} -$$
$$\frac{q}{2}\left[\left(\sqrt{\frac{q^2}{4} - 2\beta q + \beta q^2 + \beta + \gamma q} + q\sqrt{\varepsilon_n + \beta}\right)\right] \quad (24)$$

Substituting Eq. (24) into Eq. (11) and carrying out simple algebra, where

$$\tau'(s) = -2q - 2\left(\sqrt{\left(\frac{1}{4} - \delta + \gamma - \eta - \beta\right)} + q\sqrt{(\varepsilon_n + \beta)}\right) < 0 \quad (25)$$

and

$$\sigma''(s) = -2q \quad (26)$$

We have

$$\varepsilon_n = -\beta + \frac{1}{4}\left[\frac{\beta\left(\frac{1}{q^2}-1\right)-\left(n+\frac{1}{2}+\sqrt{\frac{1}{4}-\frac{2\beta}{q}+\beta+\frac{\beta}{q^2}-\frac{\gamma}{q}}\right)^2}{\left(n+\frac{1}{2}+\sqrt{\frac{1}{4}-\frac{2\beta}{q}+\beta+\frac{\beta}{q^2}-\frac{\gamma}{q}}\right)}\right]^2 \quad (27)$$

Substituting Eqs. (17) into Eq. (27) gives the energy eigenvalue equation of the Hua potential in the form

$$E_{n\ell}^{(Approx.2)} = V_1 - \frac{\hbar^2\alpha^2}{2\mu}\left[\frac{\frac{\mu V_1}{2\hbar^2\alpha^2}\left(\frac{1}{q^2}-1\right)-\zeta^2}{\zeta}\right]^2 \quad (28)$$

The corresponding wave functions can be evaluated by substituting $\pi(s)_-$ and $\sigma(s)$ from Eq. (18) and Eq. (21) respectively into Eq. (5) and solving the first order differential equation. This gives

$$\Phi(s) = s^{\sqrt{\varepsilon_n + \beta}}(1 - qs)^{\frac{1}{2}+\sqrt{\frac{1}{4}-\frac{2\beta}{q}+\beta+\frac{\beta}{q^2}+\frac{\gamma}{q}}} \quad (29)$$

The weight function $\rho(s)$ from Eq. (7) can be obtained as

$$\rho(s) = s^{2\sqrt{\varepsilon_n + \beta}}(1 - qs)^{2\sqrt{\frac{1}{4}-\frac{2\beta}{q}+\beta+\frac{\beta}{q^2}+\frac{\gamma}{q}}} \quad (30)$$

From the Rodrigues relation of Eq. (6), we obtain

$$y_n(s) \equiv N_{n,l}P_n^{\left(2\sqrt{\varepsilon_n+\beta},2\sqrt{\frac{1}{4}-\frac{2\beta}{q}+\beta+\frac{\beta}{q^2}+\frac{\gamma}{q}}\right)}(1 - 2qs) \quad (31)$$

where $P_n^{(\theta,\vartheta)}$ is the Jacobi Polynomial.

Substituting $\Phi(s)$ and $y_n(s)$ from Eq. (29) and Eq. (31) respectively into Eq. (3), we obtain the unnormalised wave function as;

$$R_n(s) = N_{n,l} s^{\sqrt{\varepsilon_n+\beta}} (1-qs)^{\frac{1}{2}+\sqrt{\frac{1}{4}-\frac{2\beta}{q}+\beta+\frac{\beta}{q^2}+\frac{\gamma}{q}}} P_n^{\left(2\sqrt{\varepsilon_n+\beta},\, 2\sqrt{\frac{1}{4}-\frac{2\beta}{q}+\beta+\frac{\beta}{q^2}+\frac{\gamma}{q}}\right)} (1-2qs) \qquad (32)$$

Again by using approximation 1 (Equation 1b) and repeating the above procedure, we can consequently obtain the energy eigenvalues as

$$E_{n\ell}^{(Approx.1)} = V_1 + \frac{2\hbar^2\alpha^2 C_0 \ell(\ell+1)}{\mu} - \frac{\hbar^2\alpha^2}{2\mu}\left[\frac{\frac{\mu V_1}{2\hbar^2\alpha^2}\left(\frac{1}{q^2}-1\right)-\zeta^2}{\zeta}\right]^2 \qquad (33)$$

$$\zeta = \left(n + \frac{1}{2} + \sqrt{\frac{1}{4} + \frac{\mu V_1}{2\hbar^2\alpha^2}\left(\frac{1}{q}-1\right)^2 + \frac{\ell(\ell+1)}{q}}\right) \qquad (34)$$

## 4. CONCLUSION

In this work, we have investigated non-relativistic problem of Schrodinger equation subject to the Hua potential respectively within the framework of two approximations scheme. We have obtained exact energy eigenvalues equation and radial wave functions using the NU method. Finally, we note that our result agree with existing literature.